\documentclass[twocolumn,superscriptaddress,showpacs,amsmath,amssymb]{revtex4}
\usepackage{}
\usepackage{amssymb}
\usepackage{amsmath}
\usepackage{amsfonts}
\usepackage{mathrsfs}

\usepackage{bm}
\usepackage{graphicx}

\usepackage{hyperref}

\newcommand{\nl}{\nonumber \\}

\newcommand{\be}{\begin{equation}}
\newcommand{\ee}{\end{equation}}
\newcommand{\bea}{\begin{eqnarray}}
\newcommand{\eea}{\end{eqnarray}}
\newcommand{\bal}{\begin{align}}
\newcommand{\eal}{\end{align}}
\newcommand{\bsube}{\begin{subequations}}
\newcommand{\esube}{\end{subequations}}
\newcommand{\Eq}[1]{Eq.\,(\ref{#1})}

\newcommand{\dg}{\dagger}
\newcommand{\la}{\langle}
\newcommand{\ra}{\rangle}

\begin{document}

\title{Precision control of charge coherence in parallel double dot systems through
 spin-orbit interaction}

\author{Jinshuang Jin} \email{jsjin@hznu.edu.cn}
\affiliation{ Department of Physics, Hangzhou Normal University,
  Hangzhou 310036, China}
\author{Matisse Wei-Yuan Tu}
\affiliation{Department of Physics and Center for Quantum
Information Science, National Cheng Kung University, Tainan 70101,
Taiwan}
\author{Nien-En Wang}
\affiliation{Department of Physics and Center for Quantum
Information Science, National Cheng Kung University, Tainan 70101,
Taiwan}
\author{Wei-Min Zhang} \email{wzhang@mail.ncku.edu.tw}
\affiliation{Department of Physics and Center for Quantum
Information Science, National Cheng Kung University, Tainan 70101,
Taiwan}
\date{\today}

\begin{abstract}

In terms of the exact quantum master equation solution for open
electronic systems, the coherent dynamics of two charge states described by
two parallel quantum dots with one fully polarized electron on either dot
is investigated in the presence of spin-orbit interaction.
We demonstrate that the double dot system
can stay in a dynamically decoherence free space.
The coherence between two double dot charge states can be precisely
manipulated through a spin-orbit coupling.
The effects of the temperature, the finite bandwidth of lead, and the energy deviations
during the coherence manipulation are also explored.

\end{abstract}

\maketitle


\section{Introduction}

Quantum information and quantum computation implementation in terms
of electron charges and spins have attracted tremendous attention for
the development of large-scale quantum information processing.
\cite{Los98120,Kan98133,Zha07165311,Cot10160502,Shi10195305,Fuj06759,Han071217,Liu10703}
Current experimental technology in nanofabrications allows to design
various nanostructures with turnable couplings and energy levels
through external gate voltages.
\cite{Hay03226804,Pet052180,Now071430,Pio08776,Nad101084}
The tunability of couplings and energy levels in quantum-dot-based
nanostructures becomes a promising technology for large-scale
solid-state quantum computing. \cite{Fuj06759,Han071217} However,
the manipulation of quantum coherence in quantum dots, although it
can be realized through a series of high-speed voltage pulses,
\cite{Hay03226804} is rather difficult for integrated precision
controls, mainly due to the inevitable decoherence arising from the
dissipative tunneling processes via the coupled electrodes.
\cite{Tu08235311}
In fact, to physically isolate a quantum device from various
contacts in nanostructures is almost impossible because of the
leakage effect induced by higher-order electron tunneling processes.

Recently, spin-orbit interaction in quantum dots has attracted much attention
both theoretically and experimentally
due to its potential roles in nano electronic devices and spintronics.
For instance, researchers have shown that the electron spins in quantum dots can be
electrically controlled by spin-orbit couplings. \cite{Deb05226803,Fli06240501,Gol06165319,Lev1119}
It has also been shown that
efficiently polarizing and manipulating the electron spin in a QD
via spin-orbit interaction \cite{Sun06235301}
plays an important role in the coherent interaction of qubits, \cite{Ste03115306,Ste04140501}
and provides a promising mechanism for coherent spin rotations. \cite{Now071430}
In this article, we propose a scheme for the
coherence control of two spin-polarized charge states with two parallel quantum dots
by using the Rashba spin-orbit coupling.
We analyze the coherence dynamics of electrons by solving
the exact master equation of the corresponding reduced density matrix.
\cite{Tu08235311,Jin10083013} The result indicates that the
parallel double dot can stay in a dynamically decoherence-free
coherence space, \cite{Xio12032107} in which the coherence phase between the two charge
states can be precisely controlled through the Rashba spin-orbit interaction. In
other words, the parallel double dot is effectively isolated from
its contacts and the coherence control can be largely simplified.
This may provide an alternative scheme for the
coherence manipulation of two charge states in parallel quantum dots,
where the tunable inter-dot coupling is not required.

\section{The parallel double dot system and its exact nonequilibrium dynamics}

\begin{figure}
\begin{center}
\centerline{\includegraphics*[width=0.4\columnwidth,angle=0]{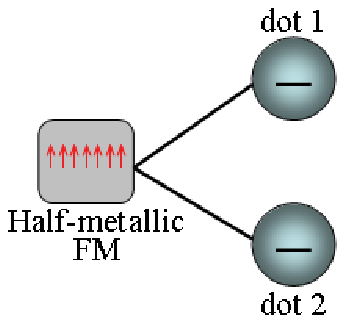}}
\end{center}
\caption{ Schematic plot of a double quantum dot coupled in parallel
with a half-metallic ferromagnetic lead. Two excess electronic
states in each dot constitutes a charge qubit. }
 \label{fig1}
\end{figure}

The nanostructure of a double quantum dot coupled in
parallel with a lead, each dot has one excess electronic state,
is shown schematically in Fig.~\ref{fig1}. The
lead should be half-metallic ferromagnetic \cite{Tom10207001} or a
mesoscopic Stern-Gerlach spin filter \cite{Ohe05041308} so that only
fully polarized spin electrons pass into each dot through the lead.
The general Hamiltonian of the nanostructure consists of three
parts:
$H = H_B+ H_S +H' $.
$H_B =\sum_{k}\epsilon_{k} c^\dag_{k}c_{k}$ is the lead Hamiltonian,
where $c^\dag_{k}, c_{k}$ are the creation and annihilation electron
operators of the lead. The double dot Hamiltonian is
$H_S=\sum_{ij}
\epsilon_{ij} a^\dag_i a_j$,
 where $a^\dag_{i} (a_i)$ creats
(destroies) an electron on the dot $i$ ($=1,2$) with the energy
$\epsilon_i=\epsilon_{ii}$, and $\epsilon_{12}=\epsilon^*_{21}$ is
the inter-dot coupling.
The coupling Hamiltonian between the dots and the lead
is given by
\be
H' = \sum_{k i}[t_{k i}
e^{i\phi_{i}}a^\dg_i c_{k} + {\rm H.c.}],
\ee
where $t_{k i}$ is a
real coupling coefficient, and $\phi_{i}$ is a phase induced
by Rashba spin-orbit (SO) interaction during the electron tunneling
betwen the lead and dots,
\cite{Sun05165310,Aha08125328,Shi10195305}
i.e., $\phi_{i}=m^\ast\alpha_i L_i/\hbar^2$ with $\alpha_i$ being the
spin-orbit interaction constant, $L_i$ the typical size
between the lead and the dot $i$, and $m^\ast$ the electron effective
mass.
Since the Rashba SO interaction only induce interlevel
(not intralevel) spin-flip couplings \cite{Sun05165310} and
each dot in the model contains only one single active energy level, there is
no spin-flip processing in our consideration. Besides, we have assumed
that the  lead is made by half-metallic ferromagnetic materials, and all electrons
in the lead are polarized in one direction. Thus, the spin index of
electrons can be dropped for simplicity.
The electron-electron interaction between the two dots has been ignored
by separating the two dots in a relatively large distance, which also makes
the inter-dot coupling almost vanishes, namely $\epsilon_{12}\simeq0$.

Quantum coherence dynamics in the above parallel double dot system
is described by the reduced density matrix $\rho(t)$ which can be
obtained by tracing over all the degrees of freedom of the lead from
the total density matrix $\rho_{\rm T}(t)$ of the nanostructure (the
double dot plus the lead), \be \rho(t)={\rm tr}_B[\rho_{\rm
T}(t)]={\rm tr}_B [e^{-iH(t-t_0)}\rho_{\rm T}(t_0)e^{iH(t-t_0)}].
\ee
This reduced density matrix can be solved from the exact master
equation we developed recently: \cite{Tu08235311,Jin10083013}
\begin{align}
\frac{d}{dt}\rho(t) =& -i[{\cal H}'_S(t),\rho(t)] + \sum_{ij}\Big\{
\gamma_{ij}^{}(t)(2a^{}_{j}\rho(t) a^{\dag}_{i}\nonumber \\
& -a^{\dag}_{i}a^{}_{j}\rho(t)-\rho(t) a^{\dag}_{i}a^{}_{j}) +
\widetilde{\gamma}^{}_{ij}(t)(a_{j}\rho(t) a^{\dag}_{i}\nonumber\\&
- a^{\dag}_{i}\rho(t) a^{}_{j} - a^{\dag}_{i}a^{}_{j}\rho(t)
+\rho(t) a^{}_{j}a^{\dag}_{i}) \Big\}\ . \label{emaster}
\end{align}
Here, ${\cal H}'_S(t)=\sum_{ij} \epsilon'_{ij}(t)a^\dag_i a^{}_j$ is
the modified Hamiltonian of the double dot, due to the coupling
between the system and the environment.
${\bm \epsilon}'$
and all other time-dependent coefficients in Eq.~(\ref{emaster}) are
given by
${\bm \epsilon}'^{}_{ij}(t) -i{\bm \gamma}^{}_{ij}(t) = i[\dot{\bm u}\bm
u^{-1}]^{}_{ij}$ and
$\widetilde{\bm \gamma}^{}_{ij}(t)=[\dot{\bm u}\bm u^{-1}\bm v + {\rm
H.c.}-\dot{\bm v}]^{}_{ij}$
where the matrix functions ${\bm u}$ and ${\bm v}$ obey the
 integrodifferential equations of motion \cite{Tu08235311}
\begin{subequations}
\label{uvn}
\begin{align}
\frac{d}{d\tau}{\bm u}(\tau )+  i\bm \epsilon & {\bm u}(\tau)  +
\int_{t_0}^{\tau } d\tau' \bm g (\tau-\tau') {\bm
u}(\tau')=0\ ,  \label{ue}\\
\frac{d}{d\tau}{\bm  v}(\tau)+  i\bm  \epsilon &{\bm  v}(\tau)  +
\int_{t_0}^{\tau } d\tau' \bm g (\tau-\tau') \bm v (\tau') \notag \\
& =
 \int_{t_0}^{t }d\tau'
\widetilde{\bm g} (\tau-\tau')\bm u^\dag(t-\tau'+t_0)\ , \label{ve}
\end{align}
\end{subequations}
with the initial conditions $\bm u(t_{0})=I, \bm v(t_0)=0$. In
Eq.~(\ref{uvn}), $\bm  \epsilon$ is the energy matrix of the central system
while ${\bm g}$ and $\tilde{\bm
g}$ are the time correlation
functions of the leads:
\begin{subequations}
\label{gtg}
\begin{align} & {\bm
g}(t-\tau)=\int\frac{d\omega}{2\pi}{\bm
\Gamma}(\omega)e^{-i\omega(t-\tau)}, \\
& \widetilde{\bm
g}(t-\tau)=\int\frac{d\omega}{2\pi}f(\omega){\bm
\Gamma}(\omega)e^{-i\omega(t-\tau)},
\end{align}
\end{subequations} and
$f(\omega)=1/[e^{\beta(\omega-\mu)}+1]$ is the
fermi distribution function for the contacted lead  with chemical
potential $\mu$ at initial temperature $\beta=1/k_{B}T$.
The spectral density $ \bm \Gamma(\omega) \equiv
\{\Gamma_{ij}(\omega)\}$ defined by
 \be\label{Gam}
  \Gamma_{ij}(\omega) =2\pi \sum_k t_{ki}t^*_{kj}
  e^{i(\phi_{i}-\phi_{j})}\delta(\omega-\epsilon_{k})
  \ee
summarizes the effects of the electron reservoirs on
the central system. In fact, Eq.~(\ref{ve}) has a general solution
in terms of $\bm u(\tau)$, \cite{Jin10083013}
\begin{align}
\bm v(t)= \int_{t_0}^{t}d\tau_{1}\int_{t_0}^{t}d\tau_{2} {\bm u}
(\tau_{1}) \widetilde{\bm g}(\tau_{2}-\tau_{1})
 u^\dag(\tau_2).\label{svp}
\end{align} Solving
Eq.(\ref{ue}) allows us to obtain the full information of the
electron transport and coherence dynamics for this parallel double
dot system.

On the other hand, from the exact master equation, we can also easily obtain the single
particle reduced density matrix of the double dot: \cite{Jin10083013}
\begin{align}\label{rho1}
\rho^{(1)}_{ij}(t)={\rm tr}_s[ a^\dag_j a_i \rho(t)] =[\bm
v(t)+\bm u(t)\bm \rho^{(1)}(t_0)\bm u^\dag(t)]_{ij},
\end{align}
Explicitly, we may denote the singly occupied states by $|1\rangle$
and $|2\rangle$ referring to occupation of the first dot and the
second dot, respectively, as the two independent charge states,
corresponding to a charge qubit in quantum information processing.
And the empty and double occupancy are denoted by
$|0\rangle$ and $|3\rangle$.
Then the reduced density matrix can be
exactly expressed as
\begin{align}
\label{rhot}
\rho(t)&=
\left(\begin{array}{cccc}
\rho_{00}(t)& 0 &0 & 0
\\
0& \rho^{(1)}_{11}(t)-\rho_{33}(t) & \rho^{(1)}_{12}(t) & 0
\\
0& \rho^{(1)}_{21}(t) & \rho^{(1)}_{22}(t)-\rho_{33}(t) & 0
\\
0 & 0 &0 & \rho_{33}(t)
\end{array}
\right),
\end{align}
in which { \bf $\rho_{ij} \equiv \la i|\rho|j\ra$}. Here $\rho_{00}(t)$ and $\rho_{33}(t)$ are respectively
the probabilities of empty and double occupation states that account
the leakage effects to the qubit system. The central $2\times 2$ block matrix  is just the density
matrix of the charge qubit in double dot nanostructures,
\cite{Fuj06759} and $\rho^{(1)}_{ij}(t)=\la a^\dag_j a_i \ra$ is the single-particle reduced density matrix
describing the probability of an electron in the state $|1\ra$ or $|2\ra$
and the electron transition between these two levels.
Using the basis of the double dots,
i.e., $\{|0\ra, |1\ra, |2\ra,|3\ra\}$, we can express the electron creation operator
in terms of a projection operator,
$a^\dg_j=|j\ra\la 0|+(-1)^{j}|3\ra\la  i|$,
where $i,j=1,2$ and $i\neq j$. Then the relation between $\rho(t)$ and
$\rho^{(1)}(t)$ given in Eq.~(\ref{rhot}) can be easily obtained.

If we prepare the double dot in the empty state $|0\rangle$ at
$t=t_{0}$, i.e., $\bm \rho^{(1)}(t_0)=0$, then the
solution of \Eq{rho1} is simply given by \Eq{svp}, \cite{Tu11115318} i.e.
\begin{align}\label{vt}
\bm \rho^{(1)}(t)=\int \frac{d\omega}{2\pi} \bm u(t,\omega)
f(\omega)\bm\Gamma(\omega)\bm u^\dag(t,\omega),
\end{align}
where $\bm u(t,\omega)=\int^t_{t_0}d\tau e^{i\omega(t-\tau)}\bm
u(t-\tau)$.
For
an initial empty dot state, we have also shown \cite{Tu11115318} that
$\rho_{00}(t)={\rm det}[I-\bm \rho^{(1)}(t)]$ and $\rho_{33}(t)={\rm
det}[\bm \rho^{(1)}(t)]$.
Thus the coherence dynamics between the two charge state is completely
determined by the single particle reduced density matrix $\bm
\rho^{(1)}(t)$ through the spectral Green function \cite{zhang12170402}
$\bm u_{ij}(t)= iG_{ij}(t,t_0) \equiv \langle \{a_i(t), a^\dag_j (t_0)\}\rangle$ of
Eq.~(\ref{ue}). Obviously, the coherence manipulation of the
charge qubit can be realized via the changes of the Fermi
distribution of the lead (see Eq.~(\ref{vt})), the inter-dot
coupling $\epsilon_{12}$ (see Eq.~(\ref{ue})), and/or the
spectral density matrix of Eq.~(\ref{Gam}).

\section{Precision control of quantum coherence through
spin-orbit interaction}

Conventionally, the coherence manipulation of a lateral double dot
is performed with the bias and gate voltages controlling the
electron tunneling between the leads and dots,
 the dot energy levels, and the
inter-dot coupling. \cite{Hay03226804,Fuj06759} However, precision
controls of the inter-dot coupling and the dot-lead potential
barriers through high-speed gate voltage pulses are rather difficult
and inevitably involve the charge noise, such as the $1/f$ noise.
Here we shall show that a precise external field control of the
quantum coherence through spin-orbit coupling in a parallel
double dot is much more reliable and simpler for coherence
manipulation in nanostructures. 
In particular, the present manipulation scheme is rather insensitive
to the charge noise \cite{Zor9613682,Sto081778,Cul09073102} with the
conditions of no inter-dot coupling at low temperature and small
bias voltage, as demonstrated below.

 According to the definition of \Eq{Gam}, the spectral density matrix elements
obey the following relations
$\Gamma_{12}(\omega) = \Gamma^\ast_{21}(\omega)$,
and
$|\Gamma_{12}(\omega)|^2=\Gamma_{11}(\omega)\Gamma_{22}(\omega)
$. We assume that the electron tunnelings between the lead and the two dots are symmetric,
namely $t_{1k}=t_{2k}$. Then $\Gamma_{11}(\omega)=\Gamma_{22}(\omega)\equiv
\Gamma(\omega)$ and $\Gamma_{12}(\omega)=\Gamma(\omega) e^{i\phi}$
where $\phi=\phi_1-\phi_2$. Thus the spectral density matrix can be expressed by
\begin{align}\label{Gamma}
\bm\Gamma(\omega)&=\Gamma(\omega) \left(\begin{array}{cc} 1&
e^{i\phi}
\\
e^{-i\phi} &1
\end{array}
\right),
\end{align}
We will show that
the spin-orbit coupling induced phase $\phi$ in the cross-correlations
of the dot-lead coupling, $\Gamma_{12/21}$,
totally determines the coherent phase between the two charge states of the
double dot without decoherence.

For the primary interest of the current experiments, \cite{Fuj06759}
we set up the double dot in degenerate: $\epsilon_1 \simeq
\epsilon_2=\epsilon_0$ and the dot-lead coupling in the wide band
limit: $\Gamma(\omega) \rightarrow \Gamma$. To avoid the electron-electron interaction
between the two dots, we also let the double dots separate in a relatively large distance.
Then the inter-dot coupling almost vanishes $\epsilon_{12} \simeq 0$.
Thus, the spectral Green function can be explicitly
solved (let $t_0=0$)
\be \bm u(t)=\frac{e^{-i\epsilon_0t}}{2}
\Big[1+e^{-\Gamma t}-\big(1-e^{-\Gamma
t}\big)(\sigma_x\cos\phi-\sigma_y\sin\phi)\Big],
\ee with
$\sigma_{x/y}$ being the Pauli matrix. The solution of the single
particle reduced density matrix can be analytically obtained from
\Eq{vt}:
\begin{align}\label{vt-result}
\bm \rho^{(1)}(t)&={\rm n}(t) \left(\begin{array}{cc} 1& e^{i\phi}
\\
e^{-i\phi} &1
\end{array}
\right),
\end{align}
where
\begin{align}\label{rmvt}
n(t)&=\int^\infty_{-\infty}\frac{d\omega}{2\pi} \frac{\Gamma}
{{\Gamma}^2+(\omega-\epsilon_0)^2}f(\omega) \nl&\quad
\times\left\{1-2e^{-{\Gamma}t}
\cos[(\omega-\epsilon_0)t]+e^{-2\Gamma t} \right\},
\end{align}
is the electron population in the dot 1, which is also the same for
the dot 2 due to the degeneracy of the double dot. It is easy to
show that the double occupation state $\rho_{33}(t)={\rm det}[\bm
\rho^{(1)}(t)]=0$.  In other words, with the initial empty state for
the degenerate double dot, the double occupation state is never
excited. Physically this is because the almost degenerate two dots
in the parallel double dot system almost decouple the whole Hilbert space
into two subspaces \cite{Xio12032107}. The doubly occupied state and the empty state
belong to different subspace. When the system is initially in the
empty state, it hardly evolves into the subspace with double occupation
state. This result is contrast to serial double quantum dots system,
where the leakage effect of double occupation is usually
neglected by hand with the assumpation of a
strong inter-dot Coulomb interaction. \cite{Shi10195305}
The full solution of the reduced density matrix of the double dot
thus becomes
\begin{align}\label{rhot-ii}
& \rho_{00}(t)=1-2{\rm n}(t), ~~ \rho_{11}(t)=\rho_{22}(t)=n(t), \nl
& \rho_{33}(t)=0, ~~ \rho_{12}(t)=\rho^\ast_{21}(t)=n(t)e^{i\phi}.
\end{align}

The above solution shows that when the double occupation state
vanishes, the two charge states (i.e. the charge qubit) are fully
determined by the single-particle reduced density matrix of
Eq.~(\ref{vt-result}) which can be rewritten in terms of a pure
state:
\begin{align}
\bm \rho_{\rm qubit}(t)&= \bm \rho^{(1)}(t) =|\psi(t)\rangle \langle
\psi(t)|
\end{align}
with $|\psi(t)\ra=\frac{c(t)}{\sqrt{2}}(|1\ra+e^{-i\phi}|2\ra)$. The
probability of the parallel double dot in such a coherent charge
state is $|c(t)|^2=2n(t)<1$ (see Eq.~(\ref{rmvt})) due to the
leakage effect given by $\rho_{00}(t)$.
While the coherence phase of the charge qubit is totally immunity
from the intrinsic
 fluctuation of the lead.
This unusual result shows that the parallel double dot can be kept
in a dynamically decoherence-free coherent space, \cite{Xio12032107} and the
coherence between the two quantum dots is totally controlled by
changing the phase $\phi$ through the spin-orbit interaction.

\begin{figure}
\centerline{\includegraphics*[width=1.\columnwidth,angle=0]{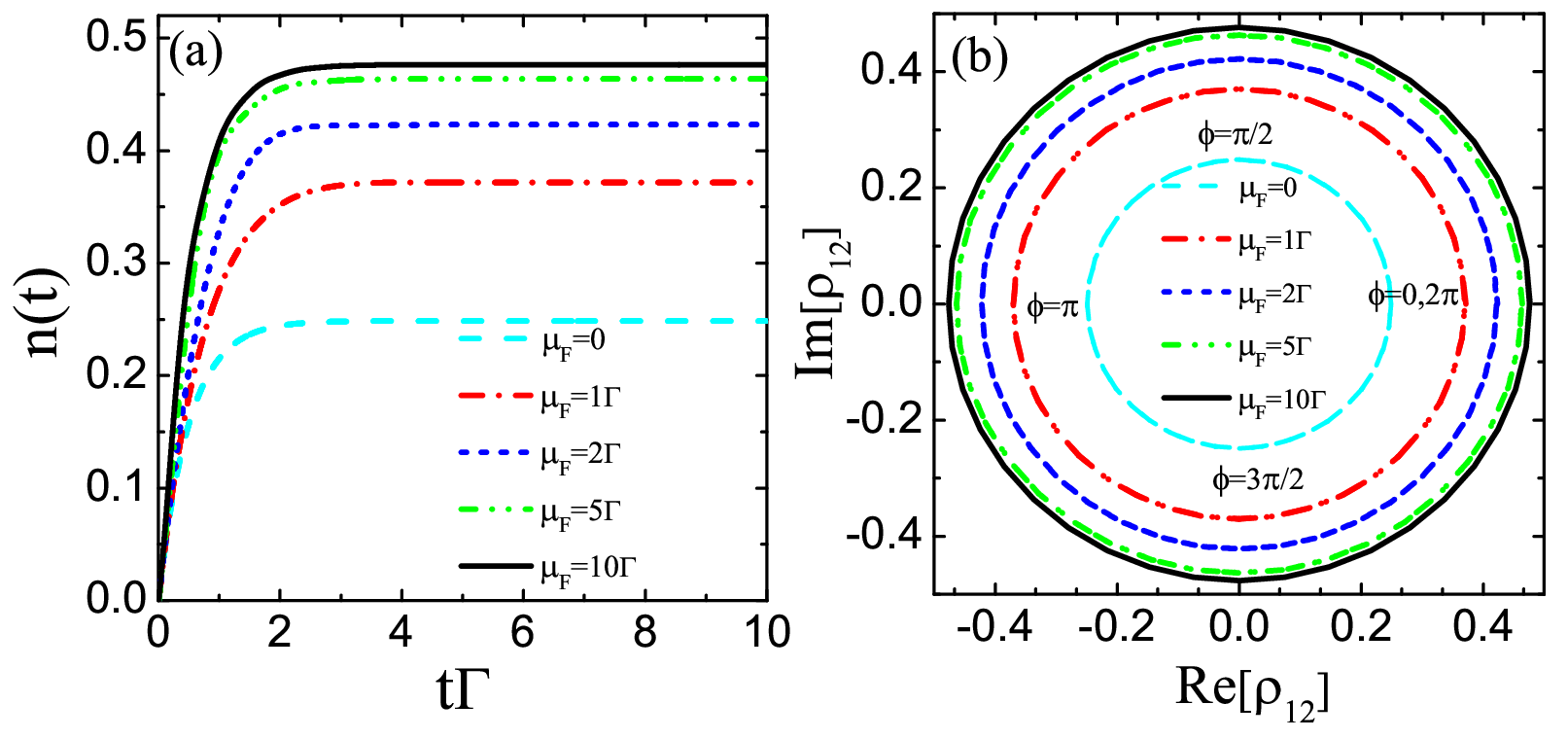}}
\caption{(a) The time evolution of electron population in the dots,
and (b) the coherence in the steady limit varying with flux $\phi$
for different bias $\mu_F=eV$. The initial temperature
$\beta^{-1}=k_{\rm B}T=0.1\Gamma$.} \label{fig2}
\end{figure}

To be more specific, we plot the electron population and the
coherence dynamics in Fig.~\ref{fig2}. It shows that $n(t)$ soon
grows to its steady values $\bar n$ within a very short time scale
($\sim 2/\Gamma$) and without further decay. The steady electron
population is given by
\begin{align}\label{rmvs}
\bar{n} &=\int^\infty_{-\infty}\frac{d\omega}{2\pi} \frac{\Gamma
f(\omega)} {{\Gamma}^2+(\omega-\epsilon_0)^2}  \simeq
\frac{1}{4}+\frac{1}{2\pi} \arctan\left(\frac{eV}
{\Gamma}\right),
\end{align}
where the second identity is for $\mu=\mu_F+\epsilon_0$ with
$\mu_F=eV$ at zero temperature. If we apply the bias to the lead
such that $eV \gg \varepsilon_0 \sim \Gamma$, then $2\bar{n} \simeq
1$, as also shown numerically in Fig.\,\ref{fig2}. Then the double
dot can stay almost in the perfect coherent state:
$|\psi\ra=\frac{1}{\sqrt{2}}(|1\ra+e^{-i\phi}|2\ra)$, where the
coherent phase $\phi$ is just the phase induced by spin-orbit
interaction
which can be tuned by applying an external electric
field.
\cite{Mat0015588,Ras03126405,Now071430,Nad101084,Fas07266801,Pfu07161308,Che091515}
Thus a decoherence-free double dot state with arbitrary coherence
phase $\phi$ can be reached via the spin-orbit interaction within a very
short time scale ($\sim 2/\Gamma$), namely a several $ps$ for a
typical quantum dot with the coupling strength of $\Gamma\sim$meV.

\begin{figure}
\centerline{\includegraphics*[width=1.0\columnwidth,angle=0]{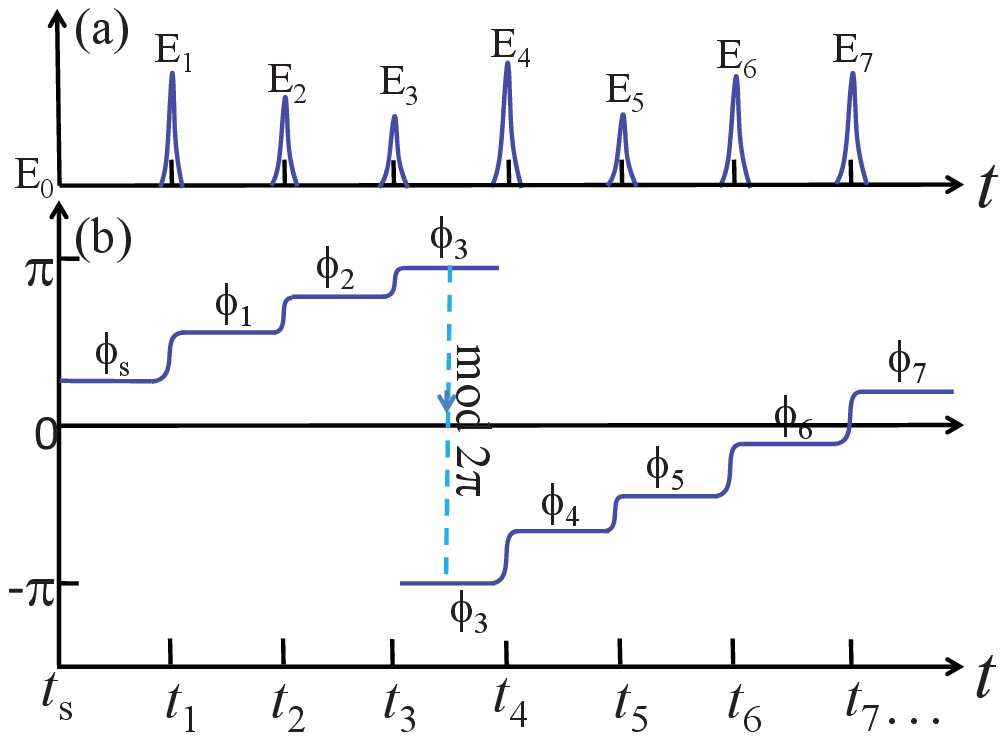}}
\caption{The coherence manipulation of the related phase
$\phi(t)=\phi_i(\text{mod}\, 2\pi)$ between the two double dot states through the
sequence temporal short pulses $E_i(t)$.} \label{fig3}
\end{figure}

To further demonstrate the precision control of coherence, one can initially
apply a constant field $E_0$ on the top of the tunnel barriers
between the lead and one dot, \cite{Nad101084} say, dot-1, and let
the spin-orbit interaction in dot-2 fixed or vanished, since the coherent phase
is just the phase difference between the two dots induced by spin-orbit coupling
as described below \Eq{Gamma}.
Then let the double
dot evolute into the steady state
$|\psi_s\ra=\frac{1}{\sqrt{2}}(|1\ra+e^{-i\phi_s}|2\ra)$ within a
time $t_s \sim 2/\Gamma$ through the Rashba SO interaction, as
we have shown above.  One can further tune the SO interaction $\alpha$
by applying a perpendicular short gate voltage pulse to manipulate the coherence
phase $\phi$. Since the double dot coherent state is
decoherence-free, it will only be instantly changed with its phase
$\phi$ just after a temporal short electric field pulse $E(t)$ is
applied. Thus, through a sequence temporal short pulses $E_j(t)$, as
plotted in Fig.\,\ref{fig3}(a), the coherence phase is changed
accordingly to the value $\phi_j$ one needed, as shown schematically
in Fig.\,\ref{fig3}(b). While the double dot maintains in the
decoherence-free state
$|\psi_j\ra=\frac{1}{\sqrt{2}}(|1\ra+e^{-i\phi_j}|2\ra)$ where only
the coherence phase is changed after the short pulse $E_j(t)$.
Taking the reported value of the SO interaction strength of
$\alpha\approx 3\times 10^{-11}$eV for InAs nanowire dots 
\cite{Fas07266801,Pfu07161308,Che091515} for an example, and the typical length
between the dots and the lead of $100$ nm with the electron
effective mass $m^\ast=0.05 m_e$, the phase change $\phi_i$ for each
single short pulse $E_i$ can range from $0$ to $\pi/2$.
That is, the coherence manipulation of the double dot state at
arbitrary time can always be precisely performed in the decoherence
free space after $t_s \sim 2/\Gamma$.

The physical picture of the coherence control via the SO interaction is
in terms of the cross-correlations $\Gamma_{12/21}$. There are two
tunneling channels in this devices, one is the electron tunneling
from $lead\rightarrow \text {dot}\, 1$ (called \emph{channel} 1) and
the other is from $lead \rightarrow \text {dot}\, 2$ (called\emph{
channel} 2).  For zero cross-correlation ($\Gamma_{12/21}=0$), the
electrons transfer through the tunneling \emph{channel} 1 is
independent from the electrons via the \emph{channel} 2. No electron
coherence between the two channels can be generated in this case.
When $\Gamma_{12/21} \neq 0$, the electrons transfer through the
tunneling \emph{channel} 1 is indistinguishable from the electrons
via the \emph{channel} 2. It is such an indistinguishable electron
tunneling induces the coherence of the two charge states in the
double dot. Thus the coherence of the double dot can be developed
and manipulated purely through the phase induced by SO coupling in
the cross-correlation between the
lead and the dots, i.e., the off-diagonal of \Eq{Gamma}.

\begin{figure*}
\centering
\includegraphics[width=0.95\textwidth,clip=true]{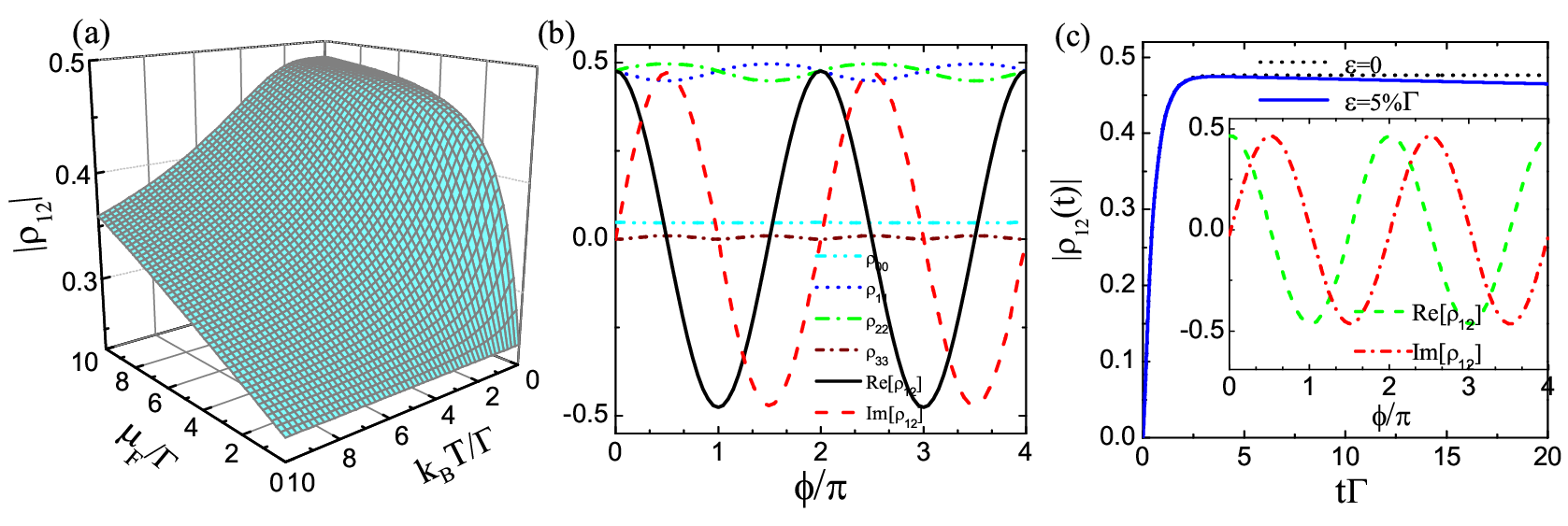}
 \caption{ (a) A 3D-plot of stationary coherent magnitude of $\rho_{12}$ with
 the temperature and the fermi surface at $\mu=\mu_F+\epsilon_0$.
 (b) The coherence dynamics of the double dot with a small inter-dot coupling, $\epsilon_{12}=2.5\%\Gamma$.
 (c) The coherence dynamics and its stationary distribution deviated away from the degenerate
 electronic states, with energy level difference $\epsilon=\epsilon_1-\epsilon_2=5\%\Gamma$.
 Other parameters $k_BT=0.1\Gamma$, $\epsilon_0=2\Gamma$, and  $\mu_F=eV=10\Gamma$.
  \label{fig4}}
\end{figure*}

Practically, the factors of thermal fluctuation, the finite
bandwidth of lead, and the deviation from the degenerate electronic
states should be considered.
Fig.\,\ref{fig4} (a) shows the
magnitude of the stationary coherence, i.e., $|\rho_{12}|=\bar{n}$
in \Eq{rmvs}, as a function of the initial temperature and the
 Fermi surfaces of the lead.
The result shows that the perfect coherence is developed with an
initially relative low temperature, as expected. Almost perfect
coherence occurs for $k_BT < 2\Gamma$, see Fig.~\ref{fig4}(a). This
can easily be realized in the current experiments, e.g., $T \simeq
350$mK for $\Gamma=30\mu$eV. \cite{Hay03226804,Hol01256802}
On the other hand, slightly splitting the degeneracy, e.g.,
$\epsilon=\epsilon_1-\epsilon_2=5\%\Gamma$, only causes a very slow
decay of the coherence, see Fig.\,\ref{fig4}(c), which could be almost
negligible. Also, a small inter-dot coupling, say
$\epsilon_{12}\simeq 2.5\% \Gamma$, does not change significantly
the coherence dynamics of the double dot, see Fog\,\ref{fig4}(b).
Regardless of the above fluctuations, the coherence phase does not
be affected.
Furthermore, we also examine the situation beyond the wide band
limit. To be specifically, we consider the energy dependence of the
spectral density of \Eq{Gamma} be a Lorentzian-type form:
\cite{Mac06085324,Jin08234703,Tu08235311}
$\Gamma(\omega)=\frac{\Gamma d^2} {(\omega-\mu)^2+d^2}$, where $d$
describes the bandwidth of the lead.
Obviously the wide band limit $d \rightarrow \infty$ leads to
$\Gamma(\omega)\rightarrow\Gamma$.
Conventionally, a finite bandwidth can induce strong non-Markovian
memory effect to the transient electron dynamics.
\cite{Jin10083013,Jin08234703,Tu08235311}
Here, when the Fermi surface of the lead is much higher than the
energy level of the double dot, the finite bandwidth effect is also
negligible, because the involved tunneling electrons have been
restricted to the region near the Fermi surface.
In addition, a much larger degenerate splitting
$\epsilon=\epsilon_1-\epsilon_2$ through the gate voltages can be
used for the charge state transition from the computational basis
($|i\rangle$, $i=1,2$) to the coherence state
$\frac{1}{\sqrt{2}}(|1\ra+e^{-i\phi}|2\ra)$.

\section{Conclusions}

In summary, we have demonstrated a novel method via the
spin-orbit interaction for the precision control
of quantum coherence in
solid-state double-dot nanostructures.
By solving the exact master equation of the double dot system, we
show that the perfect coherent charge state between the two dots,
$\frac{1}{\sqrt{2}}(|1\ra+e^{-i\phi}|2\ra)$, can be developed
without requiring a turnable inter-dot coupling. The coherence phase
$\phi$ between the two dot charge states is just the phase
in the cross-correlation of the dot-lead coupling, induced by
Rashaba SO interaction which can be tuned by a perpendicular gate voltage.
 Through investigating the effects of temperature and
finite band width of the lead, we show that the present scheme is
experimentally reliable. Also, the simplicity of precision
coherence controls in such a method may make it very promising for
the further development of large-scale quantum integrated circuits in nanostructures.

Finally, we should also point out that the present method is generic
for coherence control in nanostructures through the tunable phase of
the cross-correlation in the system-reservoir coupling, as long as
the spectral density has the form of Eq.~(\ref{Gamma}). Besides the
example of the double dot setup through the Rashba SO coupling
specified in this article, any other system which has the property
of \Eq{Gamma} with a simple tunable phase in the cross-correlation
can become a good candidate for reliable coherence control.


\acknowledgments Support from HNUEYT, the NNSF of China (10904029 and
111274085), the ZJNSF of China (Y6090345), the National Center for
Theoretical Science of ROC and the NSC of ROC
(NSC-99-2112-M-006-008-MY3) are acknowledged.


\end{document}